\documentclass[12pt]{article}
\usepackage{amsmath}
\usepackage{amssymb}
\usepackage{mathrsfs}
\usepackage{graphicx}
\usepackage{indentfirst}

\newtheorem{theorem}{Theorem}
\newtheorem{proposition}[theorem]{Proposition}
\newtheorem{lemma}[theorem]{Lemma}
\makeatletter
\def\ExtendSymbol#1#2#3#4#5{\ext@arrow 0099{\arrowfill@#1#2#3}{#4}{#5}}
\def\RightExtendSymbol#1#2#3#4#5{\ext@arrow 0359{\arrowfill@#1#2#3}{#4}{#5}}
\def\LeftExtendSymbol#1#2#3#4#5{\ext@arrow 6095{\arrowfill@#1#2#3}{#4}{#5}}
\makeatother

\begin{document}
\baselineskip 20pt

\title{Classification of the Entangled States of $2 \times L \times M \times N$}

\author{Liang-Liang Sun$^1$, Jun-Li Li$^1$ and
Cong-Feng Qiao$^{1,2}$\footnote{corresponding author;
qiaocf@ucas.ac.cn}\\[0.5cm]
\normalsize{$^{1}$School of Physics, University of Chinese Academy of
Sciences}\\
\normalsize{YuQuan Road 19A, Beijing 100049, China}\\[0.2cm]
\normalsize{$^{2}$CAS Center for Excellence in Particle Physics}\\
\normalsize{YuQuan Road 19B, Beijing 100049, China} }
\date{}
\maketitle

\begin{abstract}
We present a practical entanglement classification scheme for pure state
in form of $2\times L\times M\times N$ under the stochastic local
operation and classical communication (SLOCC), where every inequivalent
class of the entangled quantum states may be sorted out according to its
standard form and the corresponding transformation matrix. This provides a
practical method for determining the interconverting matrix between two
SLOCC equivalent entangled states, and classification examples for some
$2\times 4\times M\times N$ systems are also presented.
\end{abstract}


\section{Introduction}

Quantum theory stands as a unique pillar of physics. One of the essential
aspects providing quantum technologies an advantage over classical methods
is quantum entanglement. Quantum entanglement has practical applications in
such quantum information processing as quantum teleportation
\cite{quantum-tel}, quantum cryptography \cite{crypto-bell}, and dense
coding \cite{dense-coding92,dense-coding96}. Based on the various functions
in carrying out quantum information tasks, entanglement is classified. If
two quantum states are interconverted via stochastic local operation and
classical communication (SLOCC), they belong to the same class, and are able
to carry out the same quantum information task \cite{three-qubit}.
Mathematically, this is expressed such that the two quantum states in one
SLOCC class are connected by invertible local operators. The operator
formalism of the entanglement equivalence problem is therefore the
foundation of the qualitative and quantitative characterizations of quantum
entanglement.

Although the entanglement classification is a well-defined physical problem,
generally it is mathematically difficult, especially with the partite and
dimensions of the Hilbert space growing. Unlike the entanglement
classification under local unitary operators \cite{LU-HOSVD}, the full
classification under SLOCC for general multipartite states has solely been
obtained for up to four qubits \cite{three-qubit,four-qubit-nine}. For the
symmetric $N$-qubit state, a operational classification scheme is presented
in \cite{N-symmetric}. While in the high dimensional and less partite cases,
matrix decomposition turns out to be an effective tool for the entanglement
classification under the SLOCC \cite{2nn-onequbit}, e.g. the classification
of the $2\times M\times N$ system was completed in
\cite{2nn,2mn,2mn-parameters} and the entanglement classes of the $L\times
N\times N$ system have found to be tractable \cite{LNN}. Although an
inductive method was introduced in \cite{inductive-multi,inductive-4} to
process entangled states with more particles, its complexity substantially
grows with the increasing number of particles. By using the rank coefficient
matrices (RCM) technique \cite{coefficient-matrix}, the arbitrary
dimensional multipartite entangled states have been partitioned into
discrete entanglement families \cite{general-coefficient-matrix,222d}. As
the multipartite entanglement classes generally contain continuous
parameters which grow exponentially as the partite increases
\cite{three-qubit}, such discrete families represent a coarse grained
discrimination over the multipartite entanglement classes. Two SLOCC
inequivalent quantum states were indistinguishable when falling into the
same discrete family. Therefore, a general scheme that is able to completely
identify the different entanglement classes and determine the transformation
matrices connecting two equivalent states under SLOCC for arbitrary
dimensional four-partite states remains a significant unachieved challenge
of quantum information theory.

In this work, we present a general classification scheme for the
four-partite $2 \times L\times M\times N$ pure system, where the entangled
states are sorted into different entanglement classes under SLOCC by
utilizing the tripartite entanglement classification
\cite{2nn,2mn,2mn-parameters} and the matrix realignment technique
\cite{NKP-2000,LU-multi-kronecker}. The structure of the paper goes as
follows. In Sec. 2, the quantum states are first expressed in the
matrix-pair forms. Then the entanglement classification method is
accomplished by the construction of the standard forms from the matrix-pairs
and the determination of the transformation matrices via the matrix
realignment technique. In Sec. 3, operational considerations and some
representative examples of $2\times 4\times M\times N$ entanglement classes
are presented, where the comparison with existing results is also discussed.
Summary and conclusions are given in Sec. 4.

\section{The classification of $2\times L\times M\times N$}

\subsection{The representation of the quantum state}

A quantum state of $2\times L\times M\times N$ takes the following form
\begin{eqnarray}
|\psi\rangle = \sum_{i,l,m,n=1}^{2,L,M,N} \gamma_{ilmn} |i,l,m,n\rangle \; ,
\end{eqnarray}
where $\gamma_{ilmn} \in \mathbb{C}$ are complex numbers. In this form, the
quantum state may be represented by a high dimensional complex tensor $\psi$
whose elements are $\gamma_{ilmn}$. Two such quantum states $\psi'$ and
$\psi$ are said to be SLOCC equivalent if \cite{three-qubit}
\begin{eqnarray}
\psi' = A^{(1)} \otimes A^{(2)} \otimes A^{(3)} \otimes
 A^{(4)} \psi \; . \label{Psi-A-SLOCC}
\end{eqnarray}
Here $A^{(1)} \in \mathbb{C}^{2\times 2}$,  $A^{(2)} \in \mathbb{C}^{L\times
L}$, $A^{(3)} \in \mathbb{C}^{M\times M}$, $A^{(4)} \in \mathbb{C}^{N\times
N}$ are invertible matrices of $2\times 2$, $L\times L$, $M\times M$,
$N\times N$ separately, which act on the corresponding particles. The
transformation of the tensor elements reads
\begin{eqnarray}
\gamma\, '_{i'l'm'n'} = \sum_{i,l,m,n} A^{(1)}_{i'i} A^{(2)}_{l'l}
 A^{(3)}_{m'm} A^{(4)}_{n'n} \, \gamma_{ilmn} \; ,
\end{eqnarray}
where $\gamma'_{i'l'm'n'}$ are the tensor elements of $\psi'$, and
$A^{(k)}_{ij}$ are the matrix elements of the invertible operators
$A^{(k)}$, $k\in \{1,2,3,4\}$.

As a tensor, the quantum state $\psi$ may also be represented in the form of
a matrix-pair representation, that is $\psi \doteq
\begin{pmatrix}\Gamma_1 \\ \Gamma_2 \end{pmatrix}$. To be specific, for the
$2\times L\times M\times N$ system we have the following
\begin{eqnarray}
\psi \doteq \begin{pmatrix}\Gamma_{1} \\ \Gamma_{2}\end{pmatrix}=
\begin{pmatrix} \begin{pmatrix}
\gamma_{1111} & \gamma_{1112} & \cdots & \gamma_{11MN} \\
\gamma_{1211} & \gamma_{1212} & \cdots & \gamma_{12MN} \\
\vdots & \vdots & \ddots & \vdots \\
\gamma_{1L11} & \gamma_{1L12} & \cdots & \gamma_{1LMN}
\end{pmatrix} \\ \\
\begin{pmatrix}
\gamma_{2111} & \gamma_{2112} & \cdots & \gamma_{21MN} \\
\gamma_{2211} & \gamma_{2212} & \cdots & \gamma_{22MN} \\
\vdots & \vdots & \ddots & \vdots \\
\gamma_{2L11} & \gamma_{2L12} & \cdots & \gamma_{2LMN}
\end{pmatrix} \end{pmatrix} \; . \label{Four-Matrix-pair}
\end{eqnarray}
Here $\Gamma_{i}\in \mathbb{C}^{L\times MN}$, i.e. complex matrices of $L$
columns and $M\cdot N$ rows. For the sake of convenience, here we assume $L
< MN$ for $\Gamma_{i} \in \mathbb{C}^{L\times MN}$; while for $L \geq MN$
case, a $2\times (M\times N)\times L$ system state is represented in the
matrix-pair form of $\Gamma_i \in \mathbb{C}^{MN\times L}$. This ensures
that the matrix columns being always more than or equal to the rows.

In this matrix-pair representation, the SLOCC equivalence of two states
$\psi'$ and $\psi$ in Eq.(\ref{Psi-A-SLOCC}) transforms into the following
form
\begin{eqnarray}
\begin{pmatrix}
\Gamma_1' \\ \Gamma_2'
\end{pmatrix} = A^{(1)}
\begin{pmatrix}
P \Gamma_1 Q \\ P \Gamma_2 Q
\end{pmatrix} \; , \label{Four-Gamma}
\end{eqnarray}
where $P = A^{(2)}$, $Q^{\mathrm{T}} = A^{(3)}\otimes A^{(4)}$, $\mathrm{T}$
stands for matrix transposition, $A^{(1)}$ acts on the two matrices
$\Gamma_{1,2}$, and $P$ and $Q$ act on the rows and columns of the
$\Gamma_{1,2}$ matrices. The SLOCC equivalence of two $2\times L\times
M\times N$ quantum states in Eq.(\ref{Four-Gamma}) has a similar form as
that of the tripartite $2\times L\times MN$ pure state \cite{2mn}. The sole
difference lies in that here $Q$ is not only an invertible operator but also
a direct product of two invertible matrices, $A^{(3)}$ and $A^{(4)}$.

\subsection{Standard forms for the $2 \times L \times M\times N$ system}

The entanglement classification of the tripartite state $2\times L\times MN$
under SLOCC has already been completed in \cite{2nn,2mn}. Two tripartite
entangled states are SLOCC equivalent if and only if their standard forms
coincide. We define such standard forms of $2\times L\times MN$ to be the
standard forms of the matrix-pair of a $2\times L\times M\times N$ system,
i.e.
\begin{eqnarray}
T\otimes P\otimes Q^{\mathrm{T}}\, \psi =
T\begin{pmatrix}
P\Gamma_1 Q \\
P\Gamma_2 Q
\end{pmatrix} =
\begin{pmatrix}
E \\ J
\end{pmatrix} \; . \label{EJ-trans-matrices-tri}
\end{eqnarray}
Here $T\in \mathbb{C}^{2\times 2}$, $P\in \mathbb{C}^{L\times L}$, $Q\in
\mathbb{C}^{MN\times MN}$ are all invertible matrices, and $E$ is the unit
matrix, $J$ is in Jordan canonical form (we refer to \cite{2mn} for the
general case of the standard form). The Jordan canonical form $J$ has a
typical expression of
\begin{eqnarray}
J = \bigoplus_i J_{n_i}(\lambda_i) \; , \label{Jordan-can-form}
\end{eqnarray}
wherein $\lambda_i\in \mathbb{C}$, $J_{n_i}(\lambda_i)$ are $n_i\times n_i$
Jordan blocks
\begin{eqnarray}
J_{n_i}(\lambda_i) =
\begin{pmatrix}
\lambda_i & 1         & 0         & \cdots & 0 \\
0         & \lambda_i & 1         & \cdots & 0 \\
0         & 0         & \lambda_i & \cdots & 0 \\
\vdots    & \vdots    & \vdots    & \ddots & \vdots \\
0         & 0         & 0         & \cdots & \lambda_i
\end{pmatrix} \; .
\end{eqnarray}

For the $2\times L\times M\times N$ quantum state $\psi$ in the form of
Eq.(\ref{Four-Matrix-pair}), the following proposition exists:
\begin{proposition}
If two quantum states of\, $2\times L\times M\times N$ are SLOCC equivalent
then their corresponding matrix-pairs have the same standard forms under the
invertible operators $T\in \mathbb{C}^{2\times 2}$, $P\in
\mathbb{C}^{L\times L}$, $Q\in \mathbb{C}^{MN\times MN}$.
\label{Proposition-standard-forms}
\end{proposition}

\noindent {\bf Proof:} Suppose that two quantum states of $2\times L\times
M\times N$, $\psi$ and $\psi'$ are represented in the matrix-pairs
\begin{eqnarray}
\psi  = \begin{pmatrix} \Gamma_1  \\ \Gamma_2  \end{pmatrix} \; , \;
\psi' = \begin{pmatrix} \Gamma_1' \\ \Gamma_2' \end{pmatrix} \; .
\end{eqnarray}
The standard form of $\psi$ under the the invertible operators of $T_0 \in
\mathbb{C}^{2\times 2}$, $P_0 \in \mathbb{C}^{L\times L}$, $Q_0 \in
\mathbb{C}^{MN\times MN}$ is constructed as that of a $2\times L\times MN$
system, which is
\begin{eqnarray}
T_0 \otimes P_0 \otimes Q_0^{\mathrm{T}} \; \psi =
T_0\begin{pmatrix}
P_0 \Gamma_1 Q_0 \\
P_0 \Gamma_2 Q_0
\end{pmatrix} =
\begin{pmatrix}
E \\ J
\end{pmatrix} \; . \label{TPQ0-trans}
\end{eqnarray}
If $\psi'$ is SLOCC equivalent to $\psi$, then there exists the invertible
matrices $A^{(i)}$, such that
\begin{eqnarray}
 A^{(1)} \otimes A^{(2)}  \otimes A^{(3)}  \otimes A^{(4)} \; \psi' = \psi \; .
\end{eqnarray}
The matrix-pair form of $\psi'$ could also be transformed into
$\begin{pmatrix} E
\\ J \end{pmatrix}$ via invertible matrices, because
\begin{eqnarray}
& & T_0 A^{(1)} \otimes P_0 A^{(2)} \otimes Q_0^{\mathrm{T}} (A^{(3)} \otimes A^{(4)}) \;
\psi' \nonumber \\
& = & T_0  \otimes P_0 \otimes Q_0^{\mathrm{T}} \; \psi \nonumber \\
& = &  \begin{pmatrix} E \\ J \end{pmatrix} \; .
\end{eqnarray}
Q.E.D.

This proposition serves as a necessary condition for the SLOCC equivalence
of the entangled states of the $2\times L\times M\times N$ system. That is,
if their matrix-pair representation do not have the same standard form, the
two $2\times L\times M\times N$ entangled states are SLOCC inequivalent. The
converse of Proposition \ref{Proposition-standard-forms} is not true, which
means that different entanglement classes of $2\times L\times M\times N$
system may have the same standard form under the SLOCC .

\subsection{The transformation matrices to standard form}

The standard forms of the tripartite $2\times L\times MN$ system have been
regarded as the standard forms of the corresponding $2\times L\times M\times
N$ system, or more accurately, the entanglement families of the $2\times
L\times M\times N$ system, each of which may be transformed from entangled
states of different entanglement classes under SLOCC. In addition, the
transforming matrices $T$, $P$, $Q$ for the standard form in
Eq.(\ref{EJ-trans-matrices-tri}) were also obtained.

Generally the transformation matrices for the standard form are not unique.
For example, if $T_0$, $P_0$, $Q_0$ in Eq.(\ref{TPQ0-trans}) are the
matrices that transform $\psi$ into its standard form, then the following
matrices will do likewise
\begin{eqnarray}
T_0 \otimes SP_0 \otimes (Q_0 S^{-1})^{\mathrm{T}} \psi =
\begin{pmatrix} E \\ J \end{pmatrix} \; , \label{Matrix-pair-invariant}
\end{eqnarray}
where $SJS^{-1} = J$, i.e. $[S,J]=0$. The commutative relation implies that
if all the $\lambda_i$ in the Jordan form $J$ of Eq.(\ref{Jordan-can-form})
have geometric multiplicity 1, then the invertible matrix $S$ may be
expressed as $S = \oplus S_{n_i}$, where $S_{n_i}$ are the $n_i\times n_i$
upper triangular Toeplitz matrices conformal to the blocks of
Eq.(\ref{Jordan-can-form})
\begin{eqnarray}
S_{n_i} =
\begin{pmatrix}
s_{i1} & s_{i2}    & s_{i3}   & \cdots & s_{in_i} \\
0      & s_{i1}    & s_{i2}   & \cdots & s_{in_i-1} \\
0      & 0         & s_{i1}   & \cdots & s_{in_i-2} \\
\vdots & \vdots    & \vdots   & \ddots & \vdots \\
0      & 0         & 0        & \cdots & s_{i1}
\end{pmatrix}\; .
\end{eqnarray}
For the general case of the geometric multiplicity of $\lambda_i$, we refer
to \cite{LNN} and the references therein. There may also be an invertible
operation $S_1\in \mathbb{C}^{2\times 2}$ which acts on the first particle
and leave the ranks of the pair of matrices invariant. This operation could
be compensated by the operations on the second and third particles which
leave the standard form invariant
\begin{eqnarray}
S_1 \begin{pmatrix} S_2ES_3 \\ S_2JS_3 \end{pmatrix}
 = \begin{pmatrix} E \\ J \end{pmatrix}\; . \label{invariant-si}
\end{eqnarray}
Here the parameters in matrices $S_2\in \mathbb{C}^{L\times L}$, $S_3 \in
\mathbb{C}^{MN\times MN}$ solely depend on that of $S_1$, as shown in the
proof of the two theorems in \cite{2nn}.

\begin{figure}\centering
\scalebox{0.7}{\includegraphics{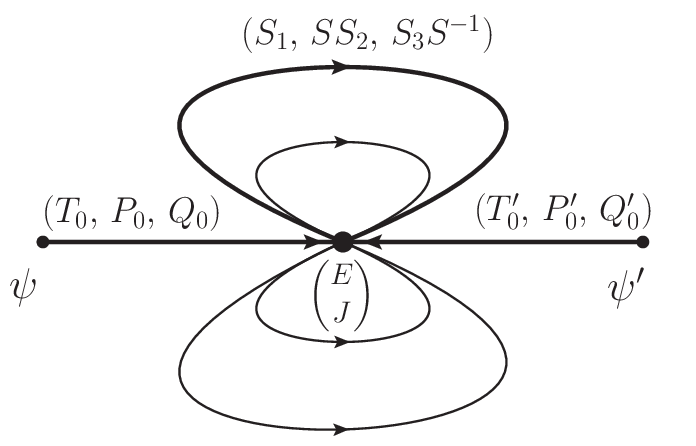}}
\caption{\small Transformation routes between two
quantum states. Two quantum states $\psi$, $\psi'$ of
$2\times L\times M\times N$ have the same standard form $(E,J)$
under the operations $(T_0,P_0,Q_0)$, and $(T_0',P_0',Q_0')$, and
$(E,J)$ is invariant under $(S_1,SS_2,S_3S^{-1})$, where all the triples of
the transformation matrices have the dimensions of
$(2\times 2, L\times L, MN\times MN)$. If there exists a route (bold line)
where $Q_0S_3S^{-1}Q_0'^{-1}$ may be written as the Kronecker product of two
invertible matrices of $\mathbb{C}^{M\times M}$ and $\mathbb{C}^{N\times N}$,
then $\psi'$ and $\psi$ are the SLOCC equivalent $2\times L\times M\times N$
entangled states.} \label{Fig-route-psi-psi'}
\end{figure}

Combining Eqs.(\ref{Matrix-pair-invariant}) and (\ref{invariant-si}), the
matrices that keep the tripartite standard forms invariant are
\begin{eqnarray}
S_1 \begin{pmatrix} SS_2 E S_3S^{-1} \\ SS_2JS_3S^{-1} \end{pmatrix}
 = \begin{pmatrix} E \\ J \end{pmatrix}\; . \label{invariant-si-s}
\end{eqnarray}
Hence, the transformation matrices which connect the two quantum states
$\psi$ and $\psi'$, which have the same standard form of matrix-pair, could
generally be written as
\begin{eqnarray}
& & T_0\otimes P_0\otimes Q_0^{\mathrm{T}} \; \psi = \begin{pmatrix} E \\ J \end{pmatrix}
= T_0'\otimes P_0'\otimes Q_0'^{\mathrm{T}} \; \psi' \nonumber \\
& \Rightarrow & \psi' = T\otimes P\otimes Q^{\mathrm{T}} \; \psi\; , \label{Four-tran-route}
\end{eqnarray}
where $T= T_0'^{-1}S_1T_0 \in \mathbb{C}^{2\times 2}$, $P=P_0'^{-1}SS_2P_0
\in \mathbb{C}^{L\times L}$, $Q^{\mathrm{T}} = Q_0S_3S^{-1}Q_0'^{-1} \in
\mathbb{C}^{MN\times MN}$, see Figure \ref{Fig-route-psi-psi'}. These
matrices may be obtained when the standard forms are constructed and their
nonuniqueness comes from the symmetries of standard forms.  A detailed
example for the construction of these matrices is presented in Sec.
\ref{example-2432}.

\subsection{The matrix realignment method} \label{sec-matrix-realignment}

To complete the entanglement classification, we introduce the matrix
realignment technique. With each matrix $A \in \mathbb{C}^{m\times n}$, the
matrix vectorization is defined to be \cite{Horn-Johnson-2}
\begin{eqnarray}
\mathrm{vec}(A) \equiv ( a_{11}, \cdots , a_{m1}, a_{12}, \cdots, a_{m2},
a_{1n}, \cdots, a_{mn} )^{\mathrm{T}} \; .
\end{eqnarray}
If the dimensions of $A$ have $m=m_1m_2$, $n=n_1n_2$, then it may be
expressed in the following block-form
\begin{eqnarray}
A = \begin{pmatrix}
A_{11} & A_{12} & \cdots & A_{1n_1} \\
A_{21} & A_{22} & \cdots & A_{2n_1} \\
\vdots & \vdots & \ddots & \vdots   \\
A_{m_11} & A_{m_12} & \cdots & A_{m_1n_1}
\end{pmatrix} \; .
\end{eqnarray}
Here $A_{ij}$ are $m_2\times n_2$ submatrices. The realignment of the matrix
$A \in \mathbb{C}^{m_1m_2\times n_1n_2}$ according to the blocks $A_{ij} \in
\mathbb{C}^{m_2\times n_2}$ is defined to be
\begin{eqnarray}
\mathscr{R}(A) \equiv
\begin{pmatrix}
\mathrm{vec}(A_{11}), \cdots, \mathrm{vec}(A_{m_11}),
\mathrm{vec}(A_{12}), \cdots, \mathrm{vec}(A_{m_12}),\cdots,\mathrm{vec}(A_{m_1n_1})
\end{pmatrix}^{\mathrm{T}} \; , \nonumber
\end{eqnarray}
where $\mathscr{R}(A) \in \mathbb{C}^{m_1n_1\times m_2n_2}$. It has been
proved that there exists a Kronecker Product singular value decomposition
(KPSVD) for the matrix $A \in \mathbb{C}^{m\times n}$ with the integer
factorizations $m=m_1m_2$ and $n=n_1n_2$, which tells \cite{NKP-2000}:
\begin{lemma}
For a matrix $A\in \mathbb{C}^{m_1m_2\times n_1n_2}$, if $\mathscr{R}(A)\in
\mathbb{C}^{m_1n_1\times m_2n_2}$ has the singular value decomposition (SVD)
$\mathscr{R}(A) = U \Sigma V^{\dag}$, where $\Sigma =
\mathrm{diag}\{\sigma_1,\cdots,\sigma_r\}$, $\sigma_i>0$ are the singular
values and $r$ is the rank of $\mathscr{R}(A)$, then $A = \sum_{k=1}^{r}
\sigma_k U_k\otimes V_k$, where $U_k \in \mathbb{C}^{m_1\times n_1}$, $V_k
\in \mathbb{C}^{m_2\times n_2}$, $\mathrm{vec}(U_k)=\sqrt{\sigma_k/\alpha_k}
\, u_k$, $\mathrm{vec}(V_k) = \sqrt{\alpha_k\sigma_k}\, v_k^*$, the scaling
parameters $\alpha_k\neq 0$ are arbitrary and $u_k$, $v_k$ are the left and
right singular vectors of $\mathscr{R}(A)$. \label{Lemma-realignment}
\end{lemma}
This technique has been applied for recognizing bipartite entanglement
\cite{Kai-Chen2003} and determining the local unitary equivalence of two
quantum states \cite{LU-multi-kronecker,LU-kroneck-decomp}. From Lemma
\ref{Lemma-realignment} we have the following corollary:
\begin{lemma}
An $MN\times MN$ invertible matrix $A$ may be expressed as the Kronecker
product of an $M\times M$ invertible matrix and an $N\times N$ invertible
matrix iff the rank of $\mathscr{R}(A)$ is 1. \label{Kroncker-decompose}
\end{lemma}

\subsection{The complete classification of the $2\times L\times M\times N$ system}

Following the preparation of Sec. \ref{sec-matrix-realignment}, the
following theorem for the entanglement classification of $2\times L\times
M\times N$ pure states under SLOCC is presented.
\begin{theorem}
Two $2\times L\times M\times N$ quantum states $\psi$ and $\psi'$ are SLOCC
equivalent iff their corresponding matrix-pair representations have the same
standard forms and the realignment of the transformation matrices $Q$ in
Eq.(\ref{Four-tran-route}) could have rank one. \label{theorem-2lmn}
\end{theorem}
\noindent {\bf Proof:} If two $2\times L\times M\times N$ quantum states
$\psi$ and $\psi'$ are SLOCC equivalent with the connecting matrices between
$\psi$ and $\psi'$ are $A^{(i)}$, $i\in \{1,2,3,4\}$
\begin{eqnarray}
\psi' = A^{(1)} \otimes A^{(2)} \otimes A^{(3)} \otimes A^{(4)} \; \psi
\; , \label{psi-psip-1}
\end{eqnarray}
then they have the same standard form in the matrix-pair form according to
Proposition \ref{Proposition-standard-forms}. Through this standard form,
there is another connecting route between $\psi$ and $\psi'$ in addition to
Eq.(\ref{psi-psip-1}), i.e.
\begin{eqnarray}
\psi' = T\otimes P \otimes Q^{\mathrm{T}} \; \psi \; . \label{psi-psip-2}
\end{eqnarray}
Combining Eq.(\ref{psi-psip-1}) and Eq.(\ref{psi-psip-2}) yields
\begin{eqnarray}
T^{-1}A^{(1)} \otimes P^{-1}A^{(2)} \otimes
((Q^{\mathrm{T}})^{-1} A^{(3)} \otimes A^{(4)})\; \psi = \psi \; .
\end{eqnarray}
As the unit matrices $E\otimes E\otimes E$ must be one of the operators
which stabilizes the quantum state $\psi$ in the matrix-pair form,
$Q^{\mathrm{T}}$ has the solution of $Q^{\mathrm{T}} = A^{(3)}\otimes
A^{(4)}$. Thus $\mathscr{R}(Q)$  could have rank one according to Lemma
\ref{Kroncker-decompose}.

If the two quantum states have the same standard forms, then we will have
Eq.(\ref{Four-tran-route}). If the matrix realignment $\mathscr{R}(Q)$
according to the factorization $MN=M\times N$ has rank one, then $Q$ may be
decomposed as $Q = Q_1\otimes Q_2$ where $Q_1\in \mathbb{C}^{M\times M}$,
$Q_2 \in \mathbb{C}^{N\times N}$. As matrix $Q$ is invertible if and only if
both $Q_1$ and $Q_2$ are invertible, thus
\begin{eqnarray}
\psi' = T\otimes P\otimes (Q_1\otimes Q_2)^{\mathrm{T}}\; \psi \; .
\end{eqnarray}
Therefore $\psi'$ and $\psi$ are SLOCC equivalent entangled states of a
$2\times L\times  M\times N$ system. Q.E.D.

To summarize, the entanglement classification scheme for the $2\times
L\times M\times N$ consists of two steps. First, the standard forms of the
matrix-pair form $2\times L\times M\times N$ quantum state $\psi$ are
constructed. By utilizing the standard forms, the entangled families of
$2\times L\times M\times N$ and the interconverting matrices between two
quantum states in the same family, $T$, $P$, $Q$, are obtained. Second, by
determining whether or not the connecting matrix $Q$ may be decomposed as
the Kronecker product of two invertible matrices via the matrix realignment
technique the SLOCC equivalence of the two quantum states is asserted. Thus
the standard form together with the route (for the connecting matrices, see
Figure \ref{Fig-route-psi-psi'}) between the quantum states form a complete
classification of the $2\times L\times M\times N$ quantum states.

\section{Examples for the entanglement classification of $2\times L\times M\times N$} \label{section-example}

\subsection{Physical considerations and the genuine entangled families}

In the field of entanglement classification, it is of great interest if we
may establish the so called operational classifications of entanglement
\cite{N-symmetric}, i.e., the different entanglement classes are related to
some experimental configuration in real physical systems. Among the possible
implementations, a two-level atom with the multimode radiation fields may be
generally considered as a system of $2\times L\times M\times N$
\cite{two-mode-JCM, El-Orany,JCMM}. This is of particular importance as the
theoretical model describing the interaction, the Jaynes-Cummings model
\cite{JC-model}, is exactly solvable and now has been extended to various
situations \cite{circuit-QED,cavity-array}.

Here we consider the genuine entanglement in the $2\times L\times M\times N$
pure system. A necessary condition for the genuine entanglement of a
$2\times L\times M\times N$ system is that all dimensions of the four
particles shall be involved in the entanglement. This requires
\begin{eqnarray}
L \leq 2MN \; , \label{genuine-dim}
\end{eqnarray}
where without loss of generality we assume the largest value of the
dimensions to be $L$. For example, a particle with dimension $25$ in a
$2\times 3\times 4\times 25$ system would always have one effective
dimension unentangled and it would have at most the genuine entanglement of
$2\times 3\times 4\times 24$. For $L = 4$, i.e. where the largest value of
the dimensions is four, the entangled systems which satisfy
Eq.(\ref{genuine-dim}) include
\begin{eqnarray}
2\times 2\times 2\times 4 \; , \;
2\times 4\times 3\times 2 \; , \; 2\times 4\times 4\times 2 \; ,\nonumber \\
2\times 4\times 3\times 3 \; , \; 2\times 4\times 4\times 3\; , \;
2\times 4\times 4\times 4  \; . \label{4-entangled-system}
\end{eqnarray}

In the construction of the standard forms (entanglement families) of
$2\times L\times M\times N$, only the operator $Q \in \mathbb{C}^{MN\times
MN}$ acts on the bipartite Hilbert spaces. As the standard forms give the
genuine entanglement of the $2\times L\times MN$ system \cite{2mn}, genuine
entanglement families of the $2\times L\times M\times N$ system are obtained
if all the dimensions of $M$ and $N$ appear in the standard forms. Therefore
the total number of such families is calculated to be
\begin{eqnarray}
\mathcal{N}_f= \sum_{i=d}^{D} \Omega_{L,i} \label{family-num}
\end{eqnarray}
where $i \in \mathbb{N}$, $d = \mathrm{max}\{M,N, \lceil L/2 \rceil\}$, $D =
\mathrm{min}\{2L,MN\}$, $\Omega_{L,i}$ are the numbers of genuine
entanglement classes of a $2\times L\times i$ system ($\Omega_{L,i}$s are
calculated from Eq.(29) in \cite{2mn-parameters}, with the class containing
parameters is counted as being one family). From Eq.(\ref{family-num}), the
numbers of entanglement families $\mathcal{N}_f$ for the systems in
Eq.(\ref{4-entangled-system}) are obtained as
\begin{eqnarray}
\mathcal{N}_f(2224) = 22 \; , \; \mathcal{N}_f(2432) = 39 \; , \;
\mathcal{N}_f(2442) = 37 \; , \nonumber \\
\mathcal{N}_f(2433) = 42 \; , \; \mathcal{N}_f(2443) = 37 \; , \;
\mathcal{N}_f(2444) = 37 \; .
\end{eqnarray}
Here $\mathcal{N}_f(2LMN)$ stands for the number genuine entanglement
families of a $2\times L\times M\times N$ system obtained from our method.

For the sake of comparison, we first list all of the entanglement families
for a $2\times 2\times (2\times 2)$ system resulting from our method. The
$\mathcal{N}_f(2222) = 5$ families includes: Two families from $2\times
2\times 2$ (GHZ and W)
\begin{eqnarray}
|\psi\rangle = |11(11)\rangle + |12(22)\rangle + |21(11)\rangle \; , \;
|\psi\rangle = |11(11)\rangle + |12(22)\rangle + |21(22)\rangle \; , \nonumber
\end{eqnarray}
two families from $2\times 2\times 3$
\begin{eqnarray}
|\psi\rangle & = & |11(11)\rangle + |12(12)\rangle + |22(21)\rangle \; , \nonumber \\
|\psi\rangle & = & |11(11)\rangle + |12(12)\rangle + |21(12)\rangle +
|22(21)\rangle \; , \nonumber
\end{eqnarray}
and one family from $2\times 2\times 4$
\begin{eqnarray}
|\psi\rangle = |11(11)\rangle + |12(12)\rangle + |21(21)\rangle +
 |22(22)\rangle \; ,\nonumber
\end{eqnarray}
where the bracket in the ket packages the particles 3 and 4.

The number of entangled families here differs from that of
\cite{four-qubit-nine} where an accidental symmetry of
$\mathrm{SU}(2)\otimes \mathrm{SU}(2)\simeq \mathrm{SO}(4)$ specific to
4-qubit states is explored, which could not be applied in more general cases
of $2\times L\times M\times N$. Within our scheme, any genuine entangled
states of $2\times 2\times 2\times 2$ system may be transformed into one of
the above standard forms (entangled families). However, according to Theorem
4, further analysis of their transformation matrices is needed in
determining the SLOCC equivalence for the two quantum states which are
assorted into the same entanglement family in our scheme. In the following,
we give examples of how our method is applied in the $2\times L\times
M\times N$ system.

\subsection{Examples of $2\times 2\times 2\times 4$}

We may package the 2 and 3 particles in the representation of the quantum
sates. The genuine entangled families of $2\times (2\times 2)\times 4$
quantum states are listed as follows. One family comes from the tripartite
$2\times 2\times 4$ system
\begin{eqnarray}
|\psi\rangle = |1(11)1\rangle + |1(22)2\rangle + |2(11)3\rangle +
 |2(22)4\rangle \; .
\end{eqnarray}
Five families come from $2\times 3\times 4$ system
\begin{eqnarray}
|\psi\rangle & = & |1(11)1\rangle + |1(12)2\rangle + |1(21)3\rangle
\hspace{3.8cm}+ |2(21)4\rangle \; , \nonumber \\
|\psi\rangle & = & |1(11)1\rangle + |1(12)2\rangle + |1(21)3\rangle + |2(11)2\rangle
\hspace{1.9cm}+ |2(21)4\rangle \; , \nonumber \\
|\psi\rangle & = & |1(11)1\rangle + |1(12)2\rangle + |1(21)3\rangle + |2(11)1\rangle
\hspace{1.9cm}+ |2(21)4\rangle \; , \nonumber \\
|\psi\rangle & = & |1(11)1\rangle + |1(12)2\rangle + |1(21)3\rangle + |2(12)3\rangle
\hspace{1.9cm}+ |2(21)4\rangle\; , \nonumber \\
|\psi\rangle & = & |1(11)1\rangle + |1(21)2\rangle + |1(21)3\rangle + |2(11)2\rangle
+ |2(12)3\rangle + |2(21)4\rangle\; . \nonumber
\end{eqnarray}
The other 16 families come from the standard forms of a $2\times 4\times 4$
system.

Therefore, there are totally 22 inequivalent families for the genuine
$2\times 2\times 2 \times 4$ entangled classes according to the present
method, while 15 distinct genuine entanglement families have been identified
in \cite{general-coefficient-matrix}. There are two merits within our
method. First, the 22 nonequivalent entanglement families correspond to a
finer grained entanglement classification under SLOCC than that of the 15
families. Second, after obtaining the entanglement families, our method also
provides a general procedure to find out the connecting matrices for two
entangled states assorted into the same family, from which the SLOCC
equivalence of the two states may be determined. While no further assessment
of equivalence of two entangled states could be made if they fall into the
same entanglement family in the coefficient matrices method.

Based on the method presented here, there also exist the continuous
entanglement families. That is, different entanglement families arise from
the different values of the characterization parameters. Here we present an
example of this kind. Among the 16 standard forms of $2\times 4\times 4$,
there is the following
\begin{eqnarray}
|\psi\rangle & = & \hspace{0.4cm} |111\rangle + \hspace{0.4cm} |122\rangle +
\hspace{0.4cm} |133\rangle + |144\rangle + \nonumber \\
& & \lambda_1 |211\rangle + \lambda_2 |222\rangle + \lambda_3 |233\rangle \; ,
\label{3lambda}
\end{eqnarray}
where  $\forall i\neq j$, $\lambda_j \neq \lambda_j$ and $\lambda_{i,j} \neq
0,1$. This corresponds to the following entanglement family of $2\times
(2\times 2)\times 4$ system
\begin{eqnarray}
|\psi\rangle & = & \hspace{0.4cm} |1(11)1\rangle + \hspace{0.4cm} |1(12)2\rangle +
\hspace{0.4cm} |1(21)3\rangle + |1(22)4\rangle + \nonumber \\
& & \lambda_1 |2(11)1\rangle + \lambda_2 |2(12)2\rangle + \lambda_3 |2(21)3\rangle \; ,
\label{4-3lambda}
\end{eqnarray}
According to the RCM method \cite{general-coefficient-matrix}, this state
would be regarded as one single family $\mathcal{F}_{4,4,4}^{\sigma_0,
\sigma_1, \sigma_2}$ regardless of the values of $\lambda_i$ (still
satisfying the condition of Eq.(\ref{3lambda})), $i\in \{1,2,3\}$, and no
further assessment of the SLOCC equivalence for the states $|\psi\rangle$
with the parameters of different values may be made. Here we show that the
state of Eq.(\ref{4-3lambda}) corresponds to a continuous entanglement
family of $2\times 2\times 2\times 4$ system according to our scheme.

First, as a $2\times 4\times 4$ state, the matrix-pair form of the state
$|\psi\rangle$ is
\begin{eqnarray}
\psi =
\begin{pmatrix}
\Gamma_1 \\
\Gamma_2
\end{pmatrix}
=
\begin{pmatrix}
E \\ J
\end{pmatrix}\; . \label{Psi-MP-244}
\end{eqnarray}
Here
\begin{eqnarray}
E = \begin{pmatrix}
1 & 0 & 0 & 0 \\
0 & 1 & 0 & 0 \\
0 & 0 & 1 & 0 \\
0 & 0 & 0 & 1
\end{pmatrix} \; , \;
J = \begin{pmatrix}
\lambda_1 & 0         & 0         & 0 \\
0         & \lambda_2 & 0         & 0 \\
0         & 0         & \lambda_3 & 0 \\
0         & 0         & 0         & 0
\end{pmatrix}\; .
\end{eqnarray}
It has already been the standard form of a $2\times 4\times 4$ system. From
\cite{2mn-parameters}, we have the following two facts concerning this
standard form. First, the operations of
\begin{eqnarray}
T =
\begin{pmatrix}
\frac{\lambda_2}{\lambda_1-\lambda_2} &
\frac{-\lambda-2}{\lambda_1(\lambda_1-\lambda_2)} \\
0 & \frac{1}{\lambda_1}
\end{pmatrix} \; , \;
P = \mathrm{diag}\{1,\frac{\lambda_1}{\lambda_2}, \frac{\lambda_1}{\lambda_3},
\frac{\lambda_1-\lambda_2}{\lambda_2}\}\; ,\; Q = E
\end{eqnarray}
will transform the state into
\begin{eqnarray}
\psi(\lambda) =
\begin{pmatrix}
\Gamma_1 \\ \Gamma_2
\end{pmatrix} =
\begin{pmatrix}
\begin{pmatrix}
0 & 0 & 0       & 0 \\
0 & 1 & 0       & 0 \\
0 & 0 & \lambda & 0 \\
0 & 0 & 0       & 1
\end{pmatrix} \\
\begin{pmatrix}
1 & 0 & 0  & 0 \\
0 & 1 & 0  & 0 \\
0 & 0 & 1  & 0 \\
0 & 0 & 0  & 0
\end{pmatrix}
\end{pmatrix} \; .
\end{eqnarray}
The continuous parameter $\lambda =
[\lambda_2(\lambda_1-\lambda_3)]/[\lambda_3(\lambda_1 - \lambda_2)]$, the
cross ratio for the quadruple $(0,\lambda_1,\lambda_2,\lambda_3)$, is
invariant under $T \in \mathbb{C}^{2\times 2}$, $P \in \mathbb{C}^{4\times
4}$ and $Q \in \mathbb{C}^{4\times 4}$ which are the invertible operators
maintaining the invariance of the standard form. Second, there is a residual
symmetry for $\lambda$ whose generators are $F(\lambda) = 1/\lambda$,
$G(\lambda) = 1-\lambda$. Thus $\psi(\lambda)$ with $\lambda \in
\mathcal{S}_{\lambda} = \{\lambda, 1/\lambda, 1-\lambda,
\lambda/(\lambda-1), 1/(1-\lambda), 1-1/\lambda \} $ are all SLOCC
equivalent. The transformation matrices for $F$ and $G$ are
\begin{eqnarray}
G & = & T\otimes P\otimes Q =
\begin{pmatrix}
-1, & 1 \\
0 & 1
\end{pmatrix} \otimes
\begin{pmatrix}
0 & 1 & 0 & 0 \\
1 & 0 & 0 & 0 \\
0 & 0 & 1 & 0 \\
0 & 0 & 0 & -1
\end{pmatrix} \otimes
\begin{pmatrix}
0 & 1 & 0 & 0 \\
1 & 0 & 0 & 0 \\
0 & 0 & 1 & 0 \\
0 & 0 & 0 & 1
\end{pmatrix} \; , \label{symmetry-G}\\
F & = & T\otimes P\otimes Q =
\begin{pmatrix}
1/\lambda, & 1 \\
0 & 1
\end{pmatrix} \otimes
\begin{pmatrix}
1 & 0 & 0 & 0 \\
0 & 0 & 1 & 0 \\
0 & 1 & 0 & 0 \\
0 & 0 & 0 & \lambda
\end{pmatrix} \otimes
\begin{pmatrix}
1 & 0 & 0 & 0 \\
0 & 0 & 1 & 0 \\
0 & 1 & 0 & 0 \\
0 & 0 & 0 & 1
\end{pmatrix} \; . \label{symmetry-F}
\end{eqnarray}
$\psi(\lambda)$ is the continuous entanglement class for $2\times 4\times 4$
system, that is different values of $\lambda$ correspond to different
entanglement classes. However when the values of $\lambda$ are in the set
$\mathcal{S}_{\lambda}$, $\psi(\lambda)$s will belong to the same
entanglement class, e.g. $\psi(2)$, $\psi(1/2)$, and $\psi(-1)$ belong to
the same $2\times 4\times 4$ entanglement class.

Now according to the scheme of theorem \ref{theorem-2lmn}, the standard
forms (the entanglement classes) of $2\times 4\times 4$ system would turn to
the entanglement families of $2\times 2\times 2\times 4$ system. Therefore
$\psi(\lambda)$ becomes the continuous entanglement family of $2\times
2\times 2\times 4$ system, where different values of $\lambda$ give rise to
different entanglement families. A subtle question arises: whether
$\psi(\lambda)$ with $\lambda \in \mathcal{S}_{\lambda}$ correspond to
different entanglement families of $2\times 2\times 2\times 4$ system or
not? To this end, we shall apply the matrix realignment method to the
transformation matrices of $P$s which connect the different states
$\psi(\lambda)$ with distinct values of $\lambda$ where $\lambda \in
\mathcal{S}_{\lambda}$. As the $P$s act on the bipartite Hilbert space of
$2\times 2$, their matrix realignment according to the factorization $4 =
2\times 2$ are
\begin{eqnarray}
\mathscr{R}(P_{G}) =
\begin{pmatrix}
0 & 1 & 1 & 0 \\
0 & 0 & 0 & 0 \\
0 & 0 & 0 & 0 \\
1 & 0 & 0 & -1
\end{pmatrix}\; , \;
\mathscr{R}(P_{F}) =
\begin{pmatrix}
1 & 0 & 0 & 0 \\
0 & 1 & 0 & 0 \\
0 & 0 & 1 & 0 \\
0 & 0 & 0 & \lambda
\end{pmatrix} \; , \label{regalign-FG}
\end{eqnarray}
where $P_{G,F}$ are just the $P$ operators that bring about the symmetry
operations $G$, $F$ in Eqs.(\ref{symmetry-G}) and (\ref{symmetry-F}). It is
clear that none of them in Eq.(\ref{regalign-FG}) can have rank one,
therefore the transformation operations relating the $\lambda$s in the set
$\mathcal{S}_{\lambda}$ cannot be decomposed into direct products of two
submatrices. We conclude that the standard forms $\psi(\lambda)$ with
different values of $\lambda$ correspond to different entanglement families,
e.g. although $\psi(2)$, $\psi(1/2)$, and $\psi(-1)$ belong to the same
entanglement class of $2\times 4\times 4$ system, but correspond to
different entanglement families of $2\times 2\times 2\times 4$ system.

\subsection{Examples of a $2\times 4 \times 3\times 2$ state} \label{example-2432}

In order to show the generalities of the method, we generate a random
quantum state for $2\times 4\times 3\times 2$ system (using built-in
function {\it RandomInteger} [1,\{4,6\}] of Mathematica), which is $\psi =
\begin{pmatrix} \Gamma_1 \\ \Gamma_2\end{pmatrix}$, where
\begin{eqnarray}
\Gamma_1 =
\begin{pmatrix}
1 & 1 & 0 & 1 & 1 & 0  \\
0 & 1 & 0 & 0 & 0 & 0  \\
0 & 1 & 0 & 0 & 0 & 1  \\
0 & 0 & 0 & 1 & 0 & 1
\end{pmatrix} \; , \;
\Gamma_2 =
\begin{pmatrix}
0 & 0 & 1 & 1 & 1 & 0  \\
1 & 0 & 1 & 1 & 0 & 1  \\
1 & 0 & 1 & 1 & 0 & 1  \\
1 & 0 & 0 & 1 & 0 & 0
\end{pmatrix} \;. \nonumber
\end{eqnarray}
In the quantum state notation, it is
\begin{eqnarray}
|\psi\rangle & = & |1111\rangle + |1112\rangle + |1122\rangle + |1131\rangle +
|1212\rangle + \nonumber \\
& & |1312\rangle + |1332\rangle + |1422\rangle + |1432\rangle + |2121\rangle +
\nonumber \\
& & |2122\rangle + |2131\rangle + |2211\rangle + |2221\rangle + |2222\rangle +
\nonumber \\
& & |2232\rangle + |2311\rangle + |2321\rangle + |2322\rangle + |2332\rangle +
\nonumber \\
& &  |2411\rangle + |2422\rangle \; .
\end{eqnarray} The rank
of $\Gamma_1$ is 4, and the following operations
\begin{eqnarray}
T_0 =
\begin{pmatrix}
1 & 0 \\
0 & 1
\end{pmatrix} \; , \;
P_0 =
\begin{pmatrix}
0 & 1  & -1 & 0 \\
1 & 2  & -3 & 2 \\
0 & -1 & 2  & -1 \\
1 & 1  & -2 & 1
\end{pmatrix} \; , \;
Q_0 =
\begin{pmatrix}
0  & -1 & 0 & 2  & 0 & -1 \\
1  & 1  & 1 & -1 & 0 & 0  \\
1  & 0  & 0 & 0  & 1 & 0  \\
0  & 1  & 0 & -1 & 0 & 0  \\
-1 & -1 & 0 & 1  & 0 & 1   \\
-1 & 0  & 0 & 0  & 0 & 0
\end{pmatrix}\; ,
\end{eqnarray}
will make
\begin{eqnarray}
\Lambda = P_0\Gamma_1 Q_0 =
\begin{pmatrix}
1 & 0 & 0 & 0 & 0 & 0  \\
0 & 1 & 0 & 0 & 0 & 0  \\
0 & 0 & 1 & 0 & 0 & 0  \\
0 & 0 & 0 & 1 & 0 & 0
\end{pmatrix} \; , \;
B = P_0 \Gamma_2 Q_0 =
\begin{pmatrix}
0 & 0 & 0 & 0 & 0 & 0  \\
0 & 0 & 0 & 1 & 0 & 0  \\
0 & 0 & 0 & 0 & 1 & 0  \\
0 & 0 & 0 & 0 & 0 & 1
\end{pmatrix} \; . \label{2432-standard}
\end{eqnarray}
The matrix-pair $\begin{pmatrix} \Lambda \\ B\end{pmatrix}$ is the standard
form for the randomly generated state $\psi$. It is invariant under the
following operations
\begin{eqnarray}
S_1 =
\begin{pmatrix}
1 & \alpha \\
0 & 1
\end{pmatrix}\; , \;
S_2 =
\begin{pmatrix}
1 & 0 & 0 & 0      \\
0 & 1 & 0 & \alpha \\
0 & 0 & 1 & 0      \\
0 & 0 & 0 & 1
\end{pmatrix}\; , \nonumber \\
S_3 = \begin{pmatrix}
1 & 0 & 0 & 0        & 0        & 0         \\
0 & 1 & 0 & -2\alpha & 0        & \alpha^2  \\
0 & 0 & 1 & 0        & -\alpha  & 0         \\
0 & 0 & 0 & 1        & 0        & -\alpha   \\
0 & 0 & 0 & 0        & 1        & 0         \\
0 & 0 & 0 & 0        & 0        & 1         \\
\end{pmatrix} \; .
\end{eqnarray}
The operations stated in Eq.(\ref{Matrix-pair-invariant}) are
\begin{eqnarray}
S =
\begin{pmatrix}
\frac{1}{a_{11}}             & 0                & 0 & 0 \\
-\frac{a_{21}}{a_{11}a_{22}} & \frac{1}{a_{22}} & 0 & 0 \\
\frac{a_{21}a_{32}- a_{22}a_{31}}{a_{11}a_{22}a_{33}} &
\frac{a_{32}}{a_{22}a_{33}} & \frac{1}{a_{33}} &
-\frac{a_{34}}{a_{22}a_{33}}\\
0 & 0 & 0 & \frac{1}{a_{22}}
\end{pmatrix}\; , \nonumber \\
S' = \begin{pmatrix}
a_{11} & 0      & 0      & 0      & 0      & 0      \\
a_{21} & a_{22} & 0      & 0      & 0      & 0      \\
a_{31} & a_{32} & a_{33} & a_{34} & 0      & 0      \\
0      & 0      & 0      & a_{22} & 0      & 0      \\
0      & 0      & 0      & a_{32} & a_{33} & a_{34} \\
0      & 0      & 0      & 0      & 0      & a_{22} \\
\end{pmatrix} \; ,
\end{eqnarray}
where $a_{ij} \in \mathbb{C}$ are arbitrary parameters which keep $S$, $S'$
invertible. The transformation matrices $(T_0,P_0,Q_0)$, and
$(S_1,SS_2,S_3S')$ are thus readily obtained from the construction of the
standard form. We refer to \cite{2mn} for the details of the construction of
the standard form of a tripartite state with one qubit.

Suppose another quantum state $\psi'$ of $2\times 4\times 3\times 2$ has the
same standard form as that of $\psi$ in Eq.(\ref{2432-standard}). We would
obtain the corresponding transformation matrices $T_0'$, $P_0'$, $Q_0'$
while constructing the standard form from $\psi'$. Thus by theorem
\ref{theorem-2lmn}, $\psi$ and $\psi'$ are SLOCC equivalent if and only if
the matrix realignment $\mathscr{R}(Q_0'^{-1}S_3S'Q_0)$ could have rank one
according to the dimensional factorization $6=2\times 3$. The example
suggests  that the scheme works better for higher dimensions, especially for
the case of $L=MN$.

\section{Conclusion}

In conclusion, we propose a practical scheme for the entanglement
classification of a $2\times L\times M\times N$ pure system under SLOCC. The
method functions by distinguishing the entanglement classes via their
standard forms together with their transformation routes to the standard
forms. Not only all the different entanglement classes but also the
transformation matrices are obtained with the method. This gives the
complete classification of the entangled states of $2\times L\times M\times
N$ under SLOCC, which has not yet been addressed in recent literature. The
method also reveals that the combination of the standard form and the routes
to the standard form may greatly reduce the complexity of the entanglement
classifications. As the entanglement generally has been considered to be the
key physical resource in quantum information science, our method may also
shed new light on the operational classifications of multipartite
entanglement with real physical systems.

\vspace{.3cm} {\bf Acknowledgments}

This work was supported in part by the National Natural Science Foundation
of China(NSFC) under grant Nos. 11121092, 11175249, 11375200 and 11205239.

\end{document}